\documentstyle[aps,twocolumn,prl,epsf]{revtex}

\begin{document}


\twocolumn[
\hsize\textwidth\columnwidth\hsize\csname @twocolumnfalse\endcsname
\title{Gap States in Dilute Magnetic Alloy Superconductors}
\author{Woonki Chung and Mark Jarrell}
\address{
	Department of Physics, University of Cincinnati, 
	Cincinnati, OH 45221-0011}
\maketitle

\widetext
\begin{abstract}
We study states in the superconducting gap induced by magnetic impurities
using self-consistent quantum Monte Carlo with
maximum entropy and formally exact analytic continuation methods.
The magnetic impurity susceptibility has different characteristics
for $T_{0} \alt T_{c0}$ and $T_{0} \agt T_{c0}$
($T_{0}$: Kondo temperature, $T_{c0}$: superconducting transition temperature)
due to the crossover between a doublet and a singlet ground state.
We systematically study the location and the weight of the gap states
and the gap parameter
as a function of $T_{0}/T_{c0}$ and the concentration of the impurities.
\end{abstract}
\pacs{74.50.+r, 75.30.Hx}

] 

\narrowtext
\paragraph*{Introduction.}
The problem of magnetic impurities in superconductors has been of
great interest for a long time.
It involves the competition of two distinct physical phenomena.
The superconducting state, characterized by a transition temperature $T_{c0}$,
is formed by the coherent pairing of electrons with time-reversal symmetry.
On the other hand, the magnetic moment on the impurity,
characterized by a Kondo temperature $T_{0}$, 
couples antiferromagnetically to the conduction electrons and 
breaks the time-reversal symmetry of the conduction electrons.
Thus even a small amount of magnetic impurities can break the pairs and 
form states within the superconducting gap.

Although many theoretical attempts~\cite{ths,ZBM,Mats,NRG,mark} 
have been made to describe these gap states induced by magnetic impurities 
in superconductors,
most of them are based upon perturbative approaches specialized for 
either $T_{0} \ll T_{c0}$ or $T_{0} \gg T_{c0}$.  
For example, Zittartz, Bringer, and M\"uller-Hartmann (ZBM)~\cite{ZBM} 
studied the problem using the Nagaoka-Suhl approximation which is a 
self-consistent high-temperature perturbation theory in the exchange 
interaction $J$ which becomes exact when $T_{0} \ll T_{c0}$.
Matsuura~\cite{Mats} used an approximate interpolation method 
between two regimes, and found that
for a single spin-$\frac{1}{2}$ impurity system the ground state (gap state) is
a singlet (doublet) for $T_{0} \gg T_{c0}$  while the ground state (gap state) 
is a doublet (singlet) for $T_{0} \ll T_{c0}$.  
It is easy to understand why the impurity has such states in these limits.  
When $T_{0}$ is vanishingly small ($J=0$) the impurity decouples from 
the superconducting host and has a doublet ground state.
Even when $T_{0}$ is finite, but small ($T_{0} \ll T_{c0}$), 
the superconductivity forms before the impurity can be screened 
by the Kondo effect,
so it maintains its doublet ground state.  However, in the limit where 
$T_{0} \gg T_{c0}$, the Kondo effect completely screens the impurity before the
superconducting transition temperature is reached.  Thus, the impurity will 
have a singlet ground state in this limit.

Nonperturbative methods have also been employed.
Satori {\it et al.}~\cite{NRG} applied 
the Numerical Renormalization Group (NRG) method to the problem of 
a single spin-$\frac{1}{2}$ impurity in a BCS superconductor at zero 
temperature.
This, however,  does not account for the self-consistent effect of a finite 
concentration of impurities or address thermodynamic properties.
Jarrell {\it et al.}~\cite{mark} introduced a self-consistent 
Quantum Monte Carlo (QMC) method to the dilute magnetic alloy superconductors
and used the Maximum Entropy Method (MEM)~\cite{MEM} to obtain 
the superconducting density of states.
Despite the fact that this QMC method treats the impurity nonperturbatively,
the MEM alone does not produce enough resolution for a systematic study
due to the complicated structure of the density of states.

In this Letter we discuss a novel method to overcome these limitations.
We use a combination of a self-consistent QMC method for the magnetic 
impurities, Eliashberg equations for the host superconductors, 
MEM  for the analytic continuation of the impurity Green's function, 
and a formally exact analytic continuation method~\cite{Mars} for 
the host Green's function. 

\paragraph*{Formalism.}
We model the magnetic impurity with a symmetric Anderson model
in the limit of infinite metallic bandwidth.
This model is characterized by an on-site Coulomb repulsion $U$
and a hybridization width $\Gamma=\pi N_0V^2$,
where $V$ is the hybridization matrix element
and $N_0$ is the density of states at the Fermi energy.
In the limit $U \gg \Gamma$, a spin-$\frac{1}{2}$ magnetic moment 
on the impurity couples antiferromagnetically to the conduction electrons 
with an exchange $J=-8\Gamma /\pi N_0U$.
Then the Kondo temperature $T_{0}$ is defined by~\cite{To}
\begin{equation}
T_{0}=U\sqrt{\frac{\Gamma}{2U}}\exp{
		\left(-\frac{\pi U}{8\Gamma}+\frac{\pi\Gamma}{2U}\right)}
\end{equation}
and it is related to $T_{K}$ defined in high-temperature perturbation theory
in $J$ by $T_{K}/T_{0} \simeq (2/\pi)^2$~\cite{Krish}.

We describe the host superconductor with the self-consistent 
Eliashberg equations.
We assume that the conduction electrons interact with phonons
with a coupling strength $\lambda(0)$ and the phonon density of states 
$F(\omega)$ consists of a Lorentzian-like peak centered at $\omega_{0}$.
With the normalization $\int_0^\infty F(\omega)d\omega=1$,
the electron-phonon spectrum function, $\alpha^2F(\omega)$, is given by
\begin{eqnarray}
\lefteqn{\alpha^2F(\omega)=\frac{\lambda(0)\omega_L(\omega_{0}^2+\omega_L^2)}
				{2\pi\omega_{0}}} \nonumber \\
 & &  \qquad \mbox{} \times \left\{
	\frac{1}{(\omega-\omega_{0})^2+\omega_L^2}-
	\frac{1}{(\omega+\omega_{0})^2+\omega_L^2} 
	\right\} \, .
\end{eqnarray}
When the width $\omega_L=0$, it represents Einstein phonon with frequency 
$\omega_{0}$. In order to have numerical stability and a continuous
density of states, we choose a finite value of $\omega_L$.

We assume that the concentration $c$ of impurities
embedded in the superconductors is small,
and there are no correlations between them.
In this dilute limit, we can describe the effect of the impurities 
on the superconducting state using a standard perturbation theory 
with an average $t$-matrix approximation.

We use $2 \times 2$ matrix Nambu formalism to represent both 
the host and impurity Green's functions.
The QMC simulation determines the fully-dressed impurity Green's function 
${\bf G}_d$ starting from the initial, $U=0$, impurity Green's function 
${\bf G}_d^0$.
Then we renormalize the host Green's function with ${\bf G}_d$, 
and update ${\bf G}_d^0$.
We iterate this self-consistent set of equations~\cite{mark} until 
convergence is reached.

This algorithm produces the fully-dressed impurity Green's function 
${\bf G}_d(\tau)$ in imaginary time as well as the gap function 
$\Delta(i\omega_n)$ and the renormalization function $Z(i\omega)$ 
in Matsubara frequencies $\omega_n=(2n+1)\pi T$.
Thus proper analytic continuations to real frequencies $\omega$
are required to obtain the density of states and the impurity spectral function.
Since it has been shown that the MEM~\cite{MEM} gives good results
for various impurity problems, we use the MEM for the impurity Green's function
to obtain the impurity spectral function ${\bf A}(\omega)$ and 
${\bf G}_d(\omega)$.
It was necessary,however, to develop a technique to analytically 
continue the non-positive definite off-diagonal term 
$({\bf A}(\omega))_{12}$, which we will discuss in a subsequent paper.
Marsiglio {\it et al.}~\cite{Mars} presented a formally exact analytic 
continuation for the Eliashberg equations.
We extend their formula to this problem by including the diagonal part
$({\bf G}_d(\omega))_{11}$ and the off-diagonal part 
$({\bf G}_d(\omega))_{12}$ of the impurity Green's function, and we have
\begin{eqnarray}
\lefteqn{
	Z(\omega ) = 1-\frac{cV^2{({\bf G}_d(\omega))}_{11}}{\omega }
	} \nonumber \\
& & \qquad \mbox{} + \frac{i\pi T}{\omega}\sum_{n=-\infty}^{\infty}
	\lambda(\omega -i\omega_n) \frac{\omega_n}
	{\sqrt{\omega_n^2+\Delta ^2(i\omega_n)}} \nonumber \\
& & \qquad \mbox{} + \frac{i\pi}{\omega} \int_{-\infty}^\infty dz \left\{ 
	\frac{\omega-z}{\sqrt{(\omega-z)^2-\Delta^2(\omega-z)}} 
		\right. \nonumber \\
& & \qquad \qquad \times \left. 
	\frac{1}{2}\alpha^2F(z)
	\left[\tanh\frac{\omega-z}{2T}+\coth\frac{z}{2T}\right] 
	\rule{0in}{4.0ex} \right\} \, , \label{eq_Z}
\end{eqnarray}
\begin{eqnarray}
\lefteqn{
	\Delta(\omega) Z(\omega) = -cV^2{({\bf G}_d(\omega))}_{12}
	} \nonumber \\
& & \qquad \mbox{} + \pi T\sum_{n=-\infty}^{\infty}
	\lambda (\omega-i\omega_n) \frac{\Delta (i\omega_n)}
	{\sqrt{\omega_n^2+\Delta^2(i\omega_n)}} \nonumber \\ 
& & \qquad \mbox{} + i\pi \int_{-\infty}^\infty dz \left\{ 
	\frac{\Delta (\omega-z)}
	{\sqrt{(\omega-z)^2-\Delta^2(\omega -z)}} \right. \nonumber \\
& & \qquad \qquad \times \left. 
	\frac{1}{2}\alpha^2F(z)
	\left[\tanh\frac{\omega-z}{2T}+\coth\frac{z}{2T} \right] 
	\rule{0in}{4.0ex} \right\}
\label{eq_phi}
\end{eqnarray}
with
\begin{equation}
\lambda (\omega ) = 
	-\int_{-\infty}^\infty d\omega' 
	\frac{\alpha ^2F(\omega' )}{\omega -\omega' +i\eta } \ .
\end{equation}
Properly taking care of the singularities in Eqs.~(\ref{eq_Z}) 
and (\ref{eq_phi})~\cite{jim},
we solve these self-consistent equations with $\Delta(i\omega_n)$ 
and ${\bf G}_d(\omega)$.
Then the superconducting density of states $N(\omega)/N_0$ is given by
\begin{equation}
\frac{N(\omega)}{N_0} = 
	\mbox{Re} \frac{\omega}{\sqrt{\omega^2-\Delta^2(\omega)}} \ .
\label{eq_nw}
\end{equation}

\begin{figure}[htb]
\epsfxsize=3.0in
\epsffile{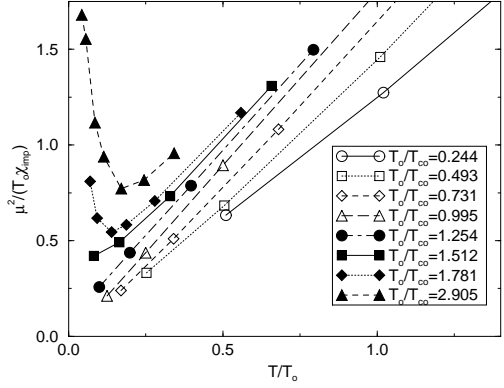}
\caption{Inverse magnetic impurity susceptibility $\chi_{\rm imp}^{-1}$ 
{\it vs.}\ $T/T_{0}$ for various values of $T_{0}/T_{c0}$. 
$\chi_{\rm imp}^{-1}$ is normalized with the magnetic moment $\mu^2$ and 
$T_{0}$.}
\label{fig1}
\end{figure}

\begin{figure}[htb]
\epsfxsize=3.0in
\epsffile{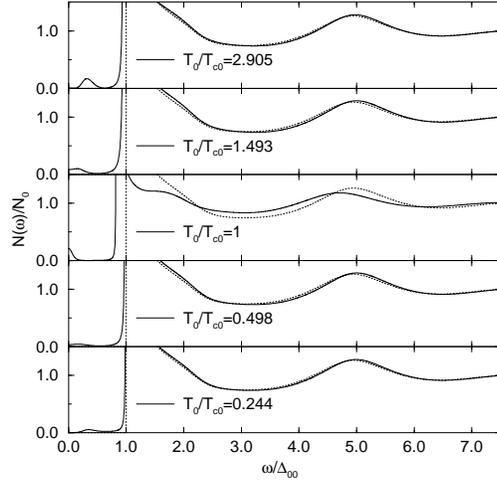}
\caption{Superconducting density of states $N(\omega)/N_0$ for various 
values of $T_{0}/T_{c0}$ when $c^*=c/[(2\pi)^2 N_0 T_{c0}]=0.03$. 
Since $N(\omega)$ is symmetric, only positive frequencies are plotted. 
The dotted line is for the pure superconductor with $T_{c0}=0.20$ 
and $T=0.05$.  $\Delta_{00}$ is the superconducting gap parameter 
for the pure superconductor.}
\label{fig2}
\end{figure}

\begin{figure}[htb]
\epsfxsize=3.0in
\epsffile{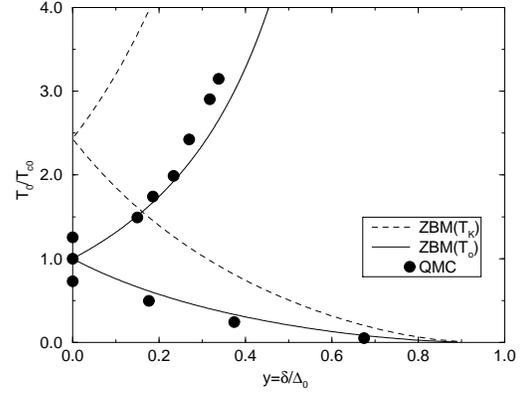}
\caption{Normalized location of gap states, $y=\delta/\Delta_{0}$, 
for various values of $T_{0}/T_{c0}$ when $c^*=c/[(2\pi)^2 N_0 T_{c0}]=0.01$. 
$\delta$ is the energy difference between the center of the gap and 
the peak of the gap states, and $\Delta_{0}$ is the superconducting gap 
parameter.  The filled circles are our QMC results.  The dashed line is the 
ZBM  result with $T_{K}$ and the solid line with $T_{0}$.  Multiple points 
appear when $y=0$ since when $T_{0} \simeq T_{c0}$ the distributions
of gap states for positive and negative $\omega$ combine into a single 
peak centered at $\omega=0$.}
\label{fig3}
\end{figure}

\begin{figure}[htb]
\epsfxsize=3.0in
\epsffile{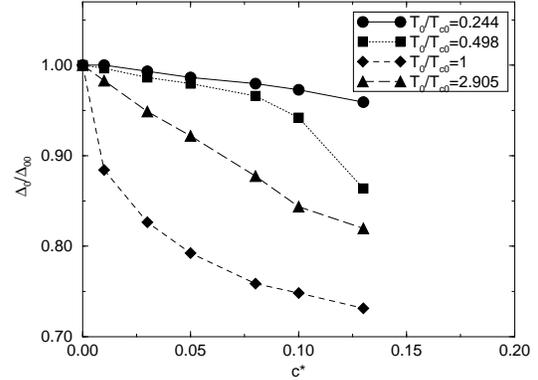}
\caption{$\Delta_{0}/\Delta_{00}$ vs.\ $c^*$ for different 
values of $T_{0}/T_{c0}$.   $\Delta_{0}$ is the superconducting gap 
parameter for the magnetic alloy superconductors and 
$\Delta_{00}$ is for the pure superconductor when $T=0.05$.}
\label{fig4}
\end{figure}

\paragraph*{Results.}
We have systematically studied the problem for many different set of parameters.
We choose $T_{c0}=0.20$ particularly for the results in
Figs.~\ref{fig1}--\ref{fig4}
such that $\omega_{0}=0.75$, $\lambda(0)=3.5$ and $\omega_L=0.25$.
First of all, we study a single impurity in a superconductor.
We may infer the form of the magnetic impurity susceptibility
$\chi_{\rm imp}$ for small $T$ based on Matsuura's observation 
regarding the crossover between a doublet and a singlet ground state, 
and we have
\begin{equation}
\lim_{T \rightarrow 0} \chi_{\rm imp}^{-1} \approx
	\left\{\begin{array}{ll}
		T & \mbox{for a doublet ground state} \\
		Te^{\delta/T} & \mbox{for a singlet ground state}
	\end{array} \right.
\label{eq_chi}
\end{equation}
where $\delta$ is the energy difference between the ground state 
and the first excited state (gap state).
Our QMC results for the inverse magnetic impurity susceptibility
$\chi_{\rm imp}^{-1}(T)$ are shown in Fig.~\ref{fig1}
for various values of $T_{0}/T_{c0}$.
With extrapolation to $T \rightarrow 0$,
$\chi_{\rm imp}^{-1}$ has zero intercepts when 
$T_{0}/T_{c0} \alt 1$ (the open symbols)
but not when $T_{0}/T_{c0} \agt 1$ (the filled symbols).
With Eq.~(\ref{eq_chi}), this implies that the doublet ground state is
favorable for $T_{0}/T_{c0} \alt 1$ and singlet for 
$T_{0}/T_{c0} \agt 1$.
It should be possible to see this behavior in NMR measurements of
$1/T_2\propto\chi_{\rm imp}$.

When the concentration of impurities is small but finite,
their effects can be seen in the dynamics of the problem.
We use the rescaled concentration of impurities
$c^*=c/[(2\pi)^2 N_0 T_{c0}]$ and choose $T=0.05$ where the superconducting
gap of the pure host is 95\% of its maximum value.
Fig.~\ref{fig2} shows the superconducting density of states
for various values of $T_{0}/T_{c0}$ when $c^*=0.03$.
For energies above the gap the phonon induced structure in the density
of states is the same as for the pure superconductor (dotted line)
but there is also structure in the gap.
As $T_{0}$ increases through $T_{c0}$ in Fig.~\ref{fig2},
the gap states move toward the center of the gap and move back to the gap edge.
The gap states form at the center of the gap at $T_{0}/T_{c0} \simeq 1$
where the crossover between doublet and singlet ground state occurs
and the pair-breaking rate is largest due to the mixing of these states.
Fig.~\ref{fig3} shows the position of the gap states
(relative to the center of the gap) for various values of $T_{0}/T_{c0}$
and the corresponding ZBM results. There is a much better fit
if we use $T_{0}$ instead of $T_{K}$ in their formula.
This is, in fact, the parameterization chosen
when their formula was fit to different experiments~\cite{Duml}.
Even so, for $T_{0}/T_{c0} \alt 1$,
the gap states move faster than ZBM's as a function of $T_{0}/T_{c0}$,
which also agrees with experiments~\cite{Duml}.

We find that the weight of the gap states (not shown here) is linearly
proportional to $c^*$, which is consistent with the experiments~\cite{Duml}
and the ZBM result.  However, while the location of the gap states is a 
universal function of $T_{0}/T_{c0}$, 
the integrated weight also depends upon the coupling $\lambda(0)$ 
(increasing with decreasing $\lambda(0)$).  
Furthermore, for fixed $\lambda(0)$, the spectral weight of the gap states 
as a function of $T_{0}/T_{c0}$ is not symmetric 
with respect to $\log^2(T_{0}/T_{c0})$ 
which is in agreement with the NRG results~\cite{NRG}.
When $T_{0}/T_{c0}\alt 1$ and the impurity ground state is a doublet, 
we find that the weight is relatively flat.  On the other hand, 
when $T_{0}/T_{c0}\agt 1$ and the impurity ground state is a singlet, 
the weight increases with $T_{0}/T_{c0}$.  

To understand the asymmetry note from Eq.~(\ref{eq_nw}) that 
any contribution to ${\rm{Im}}\Delta(\omega)$
at frequencies lower than the gap, will yield states in the gap, 
whether this contribution is from the impurity 
(i.e. $-cV^2{({\bf G}_d(\omega))}_{12}$), or 
the remaining host portion of Eq.~(\ref{eq_phi}).
At low frequencies when $T_{0}/T_{c0}<1$ both the real and imaginary parts 
of the off-diagonal impurity Green's function have opposite sign compared 
to the host result;
i.e., the real and imaginary parts of the impurity gap function 
are out of phase with the host result.  Hence the impurity and host 
contributions to the gap states tend to cancel.  
When $T_{0}/T_{c0}>1$, they have the same sign and tend to add.
Thus, when the impurity ground state crosses 
over from a doublet to a singlet, it is accompanied by a phase shift of the 
impurity order parameter~\cite{phase} and an asymmetry in the weight of
the gap states.  

As $c^*$ increases, the superconducting gap parameter $\Delta_{0}$ and the 
transition temperature $T_{c}$ for the magnetic alloy superconductors are 
reduced from the values for the pure superconductor $\Delta_{00}$ 
and $T_{c0}$ respectively.  Fig.~\ref{fig2} shows that for small $c^*$
the initial reduction of $\Delta_{0}$ is largest when $T_{0}/T_{c0} \simeq 1$.
In Fig.~\ref{fig4}, we plot $\Delta_{0}/\Delta_{00}$ {\it vs.}\ $c^*$
for different values of $T_{0}/T_{c0}$.
For larger $c^*$ the reduction of $\Delta_{0}$ is again largest for 
$T_{0}/T_{c}\simeq1$,
where $T_{c}$ is the transition temperature at that value of $c^*$.
The steepest slope of $\Delta_{0}/\Delta_{00}$ is obtained 
at larger concentration of impurities for $T_{0}/T_{c0} \alt 1$ 
and at smaller concentration of impurities for $T_{0}/T_{c0} \agt 1$. 
This may be understood as follows:
When $T_{0}/T_{c0}<1$, increasing $c^*$ decreases $T_{c}$ and thus increases
$T_{0}/T_{c}$ toward 1, and hence increases the pair breaking rate.
Therefore, when $T_{0}/T_{c0}<1$, the curvature of 
$\Delta_{0}/\Delta_{00}$ is initially negative.
In the other regime, $T_{0}/T_{c0}>1$, the initial curvature of 
$\Delta_{0}/\Delta_{00}$ is positive.

Note that the maximum initial slope
$\left| \left(\partial\Delta_{0}/\partial c\right)_{c=0} \right|$
occurs when $T_{0} \simeq T_{c0}$, but the maximum initial slope
$\left| \left(\partial T_{c}/\partial c\right)_{c=0} \right|$ occurs 
when $T_{K} \simeq T_{c0}$~\cite{markTcprl}.
These facts are independent of the coupling strength $\lambda(0)$.
This implies that $\Delta_{0}$ and $T_{c}$ have different functional forms
even to the first order in $c$; i.e., $\Delta_{0}/T_{c}$ is not constant.

\paragraph*{Conclusions.}
Using a self-consistent QMC, MEM and a formally exact analytic continuation
method~\cite{Mars}, we have calculated the density of states for a dilute
magnetic alloy superconductor.
We notice that the magnetic impurity susceptibility has different
characteristics for $T_{0}/T_{c0} \alt 1$ and $T_{0}/T_{c0} \agt 1$
due to the crossover between a doublet and a singlet impurity ground state.
The gap states form at the center of the gap when
$T_{0}/T_{c0} \mbox{ (not $T_{K}/T_{c0}$)}\simeq 1$
where the crossover between the doublet and singlet ground state occurs.
Accompanying this crossover there is a phase shift of the impurity 
order parameter, and the spectral weight of the gap states as a function 
of $T_{0}/T_{c0}$ is not symmetric with respect to $\log^2(T_{0}/T_{c0})$.
As the concentration of impurities increases, the superconducting gap parameter
$\Delta_{0}$ is reduced with different curvatures for $T_{0}/T_{c0} < 1$ and
$T_{0}/T_{c0} \agt 1$.
We also find that $\Delta_{0}/T_{c}$ for fixed $\lambda(0)$ is not constant but 
a function of both the concentration of impurities and $T_{0}/T_{c0}$ for a 
dilute magnetic alloy superconductor.

We would like to acknowledge useful discussions with
H.\ Akhlaghpour,
J.\ Deisz,
J.\ R.\ Engelbrecht,
J.\ K.\ Freericks,
B.\ Goodman, 
M.\ Ma,
P.\ Muzikar, 
and 
H.\ Pang.
WC was supported by the Office of Naval Research (ONR N00014-95-1-0883), 
and MJ by NSF (DMR 94-06678) and the Ohio Supercomputing Center.


\begin{references}
\bibitem{phase} We note that this phase shift may be observable in Josephson
tunneling through a magnetic impurity, since the impurity order parameter
vanishes when $T_{0} \simeq T_{c0}$.
\end{references}
\end{document}